\documentclass[12pt]{article}
\usepackage{a4wide,latexsym,graphicx,epsfig,psfrag,float}
\usepackage{amsmath}
\usepackage{longtable}
\usepackage{amssymb}
\usepackage{bbm}
\usepackage{cite}
\allowdisplaybreaks

\newcommand{\tppp}{\tau\to\pi\pi\pi\nu_\tau}

\def\s{\sigma}

\providecommand{\openone}{\leavevmode\hbox{\small1\kern-3.8pt\normalsize1}}

\begin{document}
\parskip=3pt plus 1pt

\begin{titlepage}
\begin{flushright}
{
FTUV/09-1123 \\
IFIC/09-54 \\
LPT-Orsay-09-91}
\end{flushright}
\vskip 1.5cm

\setcounter{footnote}{0}
\renewcommand{\thefootnote}{\fnsymbol{footnote}}

\begin{center}
{\LARGE \bf $\tau \rightarrow \pi\pi\pi\nu_\tau$ decays and the
a$_1(1260)$ off-shell width \\[10pt] revisited}
\\[50pt]

{ \sc  D.~G\'omez Dumm$^{1}$,
P.~Roig$^{2}$,
A.~Pich$^{3}$,
J. Portol\'es$^{3}$}

\vspace{1.4cm} ${}^{1)}$ IFLP, CONICET $-$ Dpto. de F\'{\i}sica, Universidad
Nacional de La Plata,  \\ C.C.\ 67, 1900 La Plata, Argentina
\\ [10pt]
${}^{2)}$ Laboratoire de Physique Th\'eorique (UMR 8627), Universit\'e de Paris-Sud XI, \\ B\^{a}timent 210,
91405 Orsay cedex, France \\[10pt]
${}^{3)}$ Departament de F\'{\i}sica Te\`orica, IFIC, CSIC ---
Universitat de Val\`encia \\
Edifici d'Instituts de Paterna, Apt. Correus 22085, E-46071
Val\`encia, Spain \\[10pt]
\end{center}
\vspace*{1cm}
\setcounter{footnote}{0}
\renewcommand{\thefootnote}{\arabic{footnote}}

\vfill

\begin{abstract}
The $\tau \rightarrow \pi \pi \pi \nu_{\tau}$ decay is driven by the
hadronization of the axial-vector current. Within the resonance chiral
theory, and considering the large-$N_C$ expansion, this process has been
studied in Ref.~\cite{GomezDumm:2003ku}. In the light of later
developments we revise here this previous work by including a new
off-shell width for the lightest a$_1$ resonance that provides a good
description of the $\tau \rightarrow \pi \pi \pi \nu_{\tau}$ spectrum and
branching ratio. We also consider the role of the $\rho(1450)$ resonance
in these observables. Thus we bring in an overall description of the $\tau
\rightarrow \pi \pi \pi \nu_{\tau}$ process in excellent agreement with
our present experimental knowledge.
\end{abstract}
\vspace*{0.5cm}
PACS~: 11.15.Pg, 12.38.-t, 12.39.Fe \\ 
Keywords~: Hadron tau decays, chiral Lagrangians, QCD, $1/N$ expansion.
\vfill

\end{titlepage}

\section{Introduction} \label{sect:1}
\hspace*{0.5cm}
The decays of the $\tau$ lepton represent an outstanding laboratory for
the analysis of various topics in particle physics. In particular, $\tau$
decays into hadrons allow to study the hadronization of vector and
axial-vector currents, thus they can be used to determine intrinsic
properties of the hadron resonances that govern the dynamics of these
processes~\cite{GomezDumm:2003ku,models,Portoles:2004vr,KS5,KS6,KS:90,Kuhn:1992nz}.
\par
At very low energies, typically $E \ll M_{\rho}$ [where $M_{\rho}$ is the
mass of the $\rho$(770) meson], chiral perturbation theory
($\chi$PT)~\cite{chpt,GL:85} is an appropriate effective theory of QCD.
However, in general this approach cannot be extended to the intermediate
energy range, in which the dynamics of resonant states plays a major role.
This is the case of hadron tau decays: these processes happen to be driven
by hadron resonances, and the corresponding energy spectrum extends over a
region where these resonances reach their on--shell peaks. In consequence,
$\chi$PT is not directly applicable to the study of the whole spectrum but
only to the very low energy domain~\cite{CFU:96}. A standard way of
dealing with these decays is to use ${\cal O}(p^2)$ $\chi$PT to fix the
normalization of the amplitudes in the low energy region, including the
effects of vector and axial--vector meson resonances by modulating the
amplitudes with {\em ad hoc} Breit--Wigner functions~\cite{KS:90,models}.
However, it has been shown that in the low energy limit this model is not
consistent with ${\cal O}(p^4)$ $\chi$PT, a fact that leads to question
the outcomes that could arise from this
procedure~\cite{GomezDumm:2003ku,Portoles:2000sr}.
\par
The significant amount of experimental data on $\tau$ decays, in
particular, $\tau\rightarrow \pi\pi\pi\nu_{\tau}$ branching ratios and
spectra~\cite{Barate:1998uf}, encourages an effort to carry out a
theoretical analysis within a model-independent framework capable to
provide information on the hadronization of the involved QCD currents. A
step in this direction has been done in Ref.~\cite{GomezDumm:2003ku},
where we have analyzed $\tau \rightarrow \pi \pi \pi \nu_{\tau}$ decays
within the resonance chiral theory
(R$\chi$T)~\cite{Ecker:1988te,Ecker:1989yg}. This procedure amounts to
build an effective Lagrangian in which resonance states are treated as
active degrees of freedom. Though the analysis in
Ref.~\cite{GomezDumm:2003ku} allows to reproduce the experimental data on
$\tau \rightarrow \pi \pi \pi \nu_{\tau}$ by fitting a few free parameters
in this effective Lagrangian, it soon would be seen that the results of
this fit are not compatible with theoretical expectations from
short-distance QCD constraints \cite{Cirigliano:2004ue}. We believe that
the inconsistency can be attributed to the usage of an ansatz for the
off-shell width of the lightest a$_1$ resonance, which was introduced {\em
ad-hoc} in Ref.~\cite{GomezDumm:2003ku}. The aim of this work is to
reanalyse $\tppp$ processes within the same general scheme, now
considering the energy-dependent width of the a$_1$ state within a
proper R$\chi$T framework.

\section{Theoretical framework} \label{sect:2}
\hspace*{0.5cm}
The construction of the effective Lagrangian in R$\chi$T
is basically ruled by the approximate chiral symmetry of QCD, which drives
the interaction of light pseudoscalar mesons, and the $SU(3)_V$
assignments of the resonance multiplets~\cite{Ecker:1988te,Ecker:1989yg}.
An additional ingredient to be taken into account is the expansion in
$1/N_C$, where $N_C$ is the number of colors in QCD~\cite{'tHooft:1973jz}.
At the leading order in this expansion one should only consider tree-level
diagrams given by a local Lagrangian with a spectrum of infinite
zero-width resonant states. However, since light resonances reach their
on--shell peaks in the energy region spanned by $\tppp$, the corresponding
resonance widths (that only appear at next-to-leading order in the
large-$N_C$ expansion) have to be also included. Moreover, as shown in
Ref.~\cite{GomezDumm:2003ku}, it is possible to obtain an adequate
description of the experimental data by including just the lowest
multiplets of vector and axial-vector resonances in the theory.
\par
We will work out $\tppp$ decays considering exact isospin symmetry. In
this limit the processes are driven only by the axial-vector current, and
appear to be dominated by the contributions of the lightest $\rho$ and
a$_1$ resonances. The corresponding effective Lagrangian in R$\chi$T
reads~:
\begin{eqnarray}
\label{eq:ret}
{\cal L}_{\rm R\chi T}   & =   &
\frac{F^2}{4}\langle u_{\mu}
u^{\mu} + \chi _+ \rangle \, + \, \frac{F_V}{2\sqrt{2}} \langle V_{\mu\nu}
f_+^{\mu\nu}\rangle \,
+ \, i \,\frac{G_V}{\sqrt{2}} \langle V_{\mu\nu} u^\mu
u^\nu\rangle  \, + \,
\frac{F_A}{2\sqrt{2}} \langle A_{\mu\nu}
f_-^{\mu\nu}\rangle \,\nonumber \\
& &  + \, {\cal L}_{\rm kin}^{\rm V} \, + \,  {\cal L}_{\rm kin}^{\rm A} \, +
\, \sum_{i=1}^{5}  \, \lambda_i  \,
{\cal O}^i_{\rm VAP} \, ,
\end{eqnarray}
where all coupling constants are real. The notation is that of
Refs.~\cite{GomezDumm:2003ku,Ecker:1988te}. Here $F$ stands for the decay
constant of the pion in the chiral limit, and the operators ${\cal
O}_{VAP}^i$ are given by~:
\begin{eqnarray}
\label{eq:lag2}
{\cal O}^1_{\rm VAP} &  = & \langle \,  [ \, V^{\mu\nu} \, , \,
A_{\mu\nu} \, ] \,  \chi_- \, \rangle \; \; , \nonumber \\
{\cal O}^2_{\rm VAP} & = & i\,\langle \, [ \, V^{\mu\nu} \, , \,
A_{\nu\alpha} \, ] \, h_\mu^{\;\alpha} \, \rangle \; \; , \\
{\cal O}^3_{\rm VAP} & = &  i \,\langle \, [ \, \nabla^\mu V_{\mu\nu} \, , \,
A^{\nu\alpha}\, ] \, u_\alpha \, \rangle \; \; ,  \nonumber \\
{\cal O}^4_{\rm VAP} & = & i\,\langle \, [ \, \nabla^\alpha V_{\mu\nu} \, , \,
A_\alpha^{\;\nu} \, ] \,  u^\mu \, \rangle \; \; , \nonumber \\
{\cal O}^5_{\rm VAP} & =  & i \,\langle \, [ \, \nabla^\alpha V_{\mu\nu} \, , \,
A^{\mu\nu} \, ] \, u_\alpha \, \rangle \nonumber \; \; .
\end{eqnarray}
Nonets of spin 1 resonances
$V$ and $A$ are described here using the antisymmetric tensor
formulation, which is consistent with the usage of the $\chi$PT Lagrangian
for light pseudoscalar mesons up to ${\cal O}(p^2)$~\cite{Ecker:1989yg}.
\par
In the Standard Model, the decay amplitudes for $\tau^- \rightarrow \pi^+
\pi^- \pi^- \nu_{\tau}$ and $\tau^- \rightarrow \pi^- \pi^0 \pi^0
\nu_{\tau}$ decays can be written as
\begin{equation}
{\cal M}_{\pm}  \, = \,  - \,
\frac{G_F}{\sqrt{2}} \, V_{ud} \, \bar u_{\nu_\tau}
\gamma^\mu\,(1-\gamma_5) u_\tau\, T_{\pm \mu} \; ,
\end{equation}
where $V_{ud}\simeq \cos\theta_C$ is an element of the
Cabibbo-Kobayashi-Maskawa matrix, and $T_{\pm\mu}$ is the hadron matrix
element of the axial-vector QCD current $A_{\mu}$,
\begin{equation} \label{eq:tmu1}
T_{\pm \mu}(p_1,p_2,p_3) \,  =  \,
 \langle  \pi_1(p_1)\pi_2(p_2)\pi^{\pm}(p_3)  |\, A_\mu\,
 e^{i {\cal L}_{QCD}} | 0  \rangle \, ,
\end{equation}
as there is no contribution of the vector current to these processes in
the isospin limit. Outgoing states $\pi_{1,2}$ correspond here to $\pi^-$
and $\pi^0$ for upper and lower signs in $T_{\pm\mu}$, respectively. The
hadron tensor can be written in terms of three form factors, $F_1$, $F_2$
and $F_P$, as \cite{Kuhn:1992nz}~:
\begin{equation}
T^{\mu} \; = \; V_1^\mu\,F_1 \, + \,  V_2^\mu\,F_2 \, + \,
Q^\mu\,F_P \; \; ,
\label{tmu}
\end{equation}
where
\begin{eqnarray}
V_1^\mu & = & \left( \, g^{\mu \nu} \, - \,
\frac{Q^{\mu} Q^{\nu}}{Q^2} \, \right) \, ( \, p_1 - p_3 \, )_{\nu} \; \; ,
\nonumber \\
V_2^\mu & = & \left( \, g^{\mu \nu} \, - \,
\frac{Q^{\mu} Q^{\nu}}{Q^2} \, \right) \, ( \, p_2 - p_3 \, )_{\nu} \; \; ,
\nonumber \\
Q^\mu & = &  p_1^\mu + p_2^\mu + p_3^\mu\; \; .
\end{eqnarray}
In this way, the terms involving the form factors $F_1$ and $F_2$ have a
transverse structure in the total hadron momenta $Q_{\mu}$, and drive a
$J^P=1^+$ transition. Meanwhile $F_P$ accounts for a $J^P=0^-$ transition
that carries pseudoscalar degrees of freedom and vanishes with the square
of the pion mass. Its contribution to the spectral function of $\tppp$
goes like $m_{\pi}^4/Q^4$ and, accordingly, it is very much suppressed
with respect to those coming from $F_1$ and $F_2$. We will not consider it
in the following.
\begin{figure}[t]
\begin{center}
\vspace*{0.9cm}
\includegraphics[scale=0.77,angle=0]{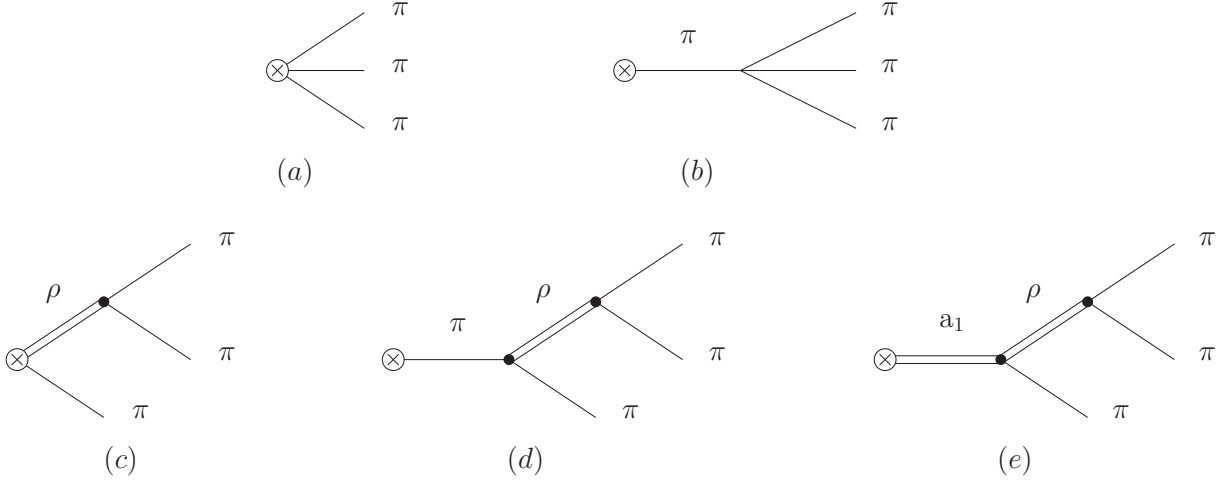}
\caption[]{\label{fig:1} \small{Diagrams contributing to the hadron
axial-vector form factors $F_i$~: (a) and (b) contribute to $F_1^{\chi}$,
(c) and (d) to $F_1^R$ and (e) to $F_1^{RR}$.}}
\end{center}
\end{figure}
\par
The evaluation of the form factors $F_1$ and $F_2$ within in the context of
R$\chi$T has been carried out in Ref.~\cite{GomezDumm:2003ku}. One has~:
\begin{equation}
 F_{\pm i} \ = \ \pm \left(
 F_i^{\chi} \, + \, F_i^{\mbox{\tiny R}} \, + \, F_i^{\mbox{\tiny RR}}\right)
\ ,\qquad i=1,2\ ,
\end{equation}
where the different contributions correspond to the diagrams in
Fig.~\ref{fig:1}. In terms of the Lorentz invariants $Q^2$,
$s=(p_1+p_3)^2$, $t=(p_2+p_3)^2$ and $u=(p_1+p_2)^2$ (notice that $u =
Q^2-s-t+3m_\pi^2$) these contributions are given
by~\cite{GomezDumm:2003ku}
\begin{eqnarray}
\label{eq:t1r}
F_1^{\chi}(Q^2,s,t) & = & - \frac{2\sqrt{2}}{3 F} \nonumber \\
F_1^{\mbox{\tiny R}}(Q^2,s,t) & = & \frac{\sqrt{2}\,F_V\,G_V}{3\,F^3} \left[
\, \frac{3\,s}{s-M_V^2} \, - \,
\left( \frac{2 G_V}{F_V} - 1 \right) \, \left(
\, \frac{2 Q^2-2s-u}{s-M_V^2} \, + \, \frac{u-s}{t-M_V^2} \,
\right)\right] \;\; \nonumber \\
F_1^{\mbox{\tiny RR}}(Q^2,s,t) & = & \frac{4 \, F_A \, G_V}{3 \,F^3} \,
 \frac{Q^2}{Q^2-M_A^2} \, \bigg[- \, (\lambda' + \lambda'')
 \, \frac{3\,s}{s-M_V^2} \, \nonumber \\
 & & \qquad\qquad\qquad\qquad\ \ + \, \, H(Q^2,s) \, \frac{2 Q^2 + s -
u}{s-M_V^2} \, + \,  H(Q^2,t) \, \frac{u-s}{t-M_V^2}\bigg] \ ,
\end{eqnarray}
where
\begin{equation} \label{eq:fq2}
H(Q^2,x)  =  - \,\lambda_0\, \frac{m_\pi^2}{Q^2} \, +  \,
\lambda'\, \frac{x}{Q^2} \, + \,  \lambda''  \; ,
\end{equation}
$\lambda_0$, $\lambda'$ and $\lambda''$ being linear combinations of the
$\lambda_i$ couplings in Eq.~(\ref{eq:ret}) that can be read in
Ref.~\cite{GomezDumm:2003ku}. Bose symmetry under the exchange of the two
identical pions in the final state implies that the form factors $F_1$ and
$F_2$ are related by $F_2(Q^2,s,t) = F_1(Q^2,t,s)$.
\par
Besides the pion decay constant $F$, the above results for the form
factors $F_i$ depend on six combinations of the coupling constants in the
Lagrangian ${\cal L}_{\rm R \chi T}$, namely $F_V$, $F_A$, $G_V$,
$\lambda_0$, $\lambda'$ and $\lambda''$ and the masses $M_V$, $M_A$ of the
vector and axial-vector nonets. All of them are in principle unknown
parameters. However, it is clear that ${\cal L}_{\rm R \chi T}$ does not
represent an effective theory of QCD for arbitrary values of the
couplings. Though the determination of the effective parameters from the
underlying theory is still an open problem, one can get information on the
couplings by assuming that the resonance region
---even when one does not include the full phenomenological spectrum---
provides a bridge between the chiral and perturbative
regimes~\cite{Ecker:1989yg}. This is implemented by matching the high
energy behaviour of Green functions (or related form factors) evaluated
within the resonance theory with asymptotic results obtained in
perturbative QCD~\cite{Moussallam:1997xx,
Knecht:2001xc,RuizFemenia:2003hm,Cirigliano:2004ue,Cirigliano:2005xn,
Mateu:2007tr,Ecker:1989yg,Amoros:2001gf}. In the $N_C \rightarrow \infty$
limit, and within the approximation of only one nonet of vector and
axial-vector resonances, the analysis of the two-point Green functions
$\Pi_{V,A}(q^2)$ and the three-point Green function VAP of QCD currents
with only one multiplet of vector and axial-vector resonances
lead to the following constraints \cite{Pich:2002xy}~:
\par
\begin{itemize}
\item[i)] By demanding that the two-pion vector form factor vanishes at
high momentum transfer one obtains the condition $F_V \, G_V =
F^2$~\cite{Ecker:1989yg}.
\item[ii)] The first Weinberg sum rule~\cite{Weinberg:1967kj} leads to
$F_V^2 - F_A^2 = F^2$, and the second Weinberg sum rule gives $F_V^2 \,
M_V^2 \, = \, F_A^2 \, M_A^2$~\cite{Ecker:1988te}.
\item[iii)] The analysis of the VAP Green function~\cite{Cirigliano:2004ue}
gives for the coupling combinations $\lambda_0$, $\lambda'$ and
$\lambda''$ entering the form factors in Eq.~(\ref{eq:t1r}) the following
results~:
\begin{eqnarray}
 \lambda' & = & \frac{F^2}{2 \, \sqrt{2} \, F_A \, G_V} \; = \;
\frac{M_A}{2 \, \sqrt{2} \, M_V} \,, \label{eq:lam1} \\[3.5mm]
\lambda'' & = & \frac{2 \, G_V \, - F_V}{2 \, \sqrt{2} \, F_A} \; = \;
\frac{M_A^2 - 2 M_V^2}{2 \, \sqrt{2} \, M_V \, M_A} \, , \label{eq:lam2} \\[3.5mm]
4 \, \lambda_0 & = & \lambda' + \lambda'' \; = \; \frac{M_A^2-M_V^2}{\sqrt{2} \, M_V \, M_A} \, , \label{eq:lam3}
\end{eqnarray}
where the second equalities in Eqs.~(\ref{eq:lam1}) and (\ref{eq:lam2})
are obtained using the above relations i) and ii).
\end{itemize}
As mentioned above, $M_V$ and $M_A$ stand for the masses of the vector
and axial-vector resonance nonets, in the chiral and large-$N_C$ limits. A
phenomenological analysis carried out in this limit~\cite{Mateu:2007tr}
shows that $M_V$ is well approximated by the $\rho(770)$ mass, whereas for
the axial mass one gets $M_{{\rm a}_1}^{1/N_C} \equiv M_A = 998 (49)$~MeV (which
differs appreciably from the presently accepted value of $M_{{\rm a}_1}$).
\par
In addition, one can require that the $J=1$ axial spectral function in
$\tppp$ vanishes for large momentum transfer. This can be seen from the
asymptotic behaviour of the axial-vector current correlator
$\Pi_A(Q^2)$~\cite{Floratos:1978jb}, taking into account that each
intermediate state carrying the appropriate quantum numbers yields a
positive contribution to Im$\Pi_A (Q^2)$. In fact, it is found that this
constraint leads to the relations in Eqs.~(\ref{eq:lam1}) and (\ref{eq:lam2}),
showing the consistency of the procedure.
\par
The above constraints allow in principle to fix all six free parameters
entering the form factors $F_i$ in terms of the vector and axial vector
masses $M_V$, $M_A$. However the
form factors in Eq.~(\ref{eq:t1r}) include zero--width $\rho$ and
a$_1$ propagator poles, which lead to divergent phase--space integrals
in the calculation of $\tppp$ decay widths. As stated above, in order to
regularize the integrals one should take into account the inclusion of
resonance widths, which means to go beyond the leading order in the $1/N_C$
expansion. In order to account for the inclusion of NLO corrections we perform the
substitutions~:
\begin{equation}
 \frac{1}{M_{R_j}^2-q^2} \; \; \longrightarrow \;\; \frac{1}{M_{j}^2-q^2- \, i \,
M_{j} \, \Gamma_{j}(q^2)} \; ,
\label{eq:subst}
\end{equation}
Here $R_j=V,A$, while the subindex $j=\rho,{\rm a}_1$ on the right hand side
stands for the corresponding physical state.
\par
The substitution in Eq.~(\ref{eq:subst}) implies the introduction of
additional theoretical inputs, in particular, the behaviour of resonance
widths off the mass shell, to which now we turn.

\section{Energy-dependent widths of resonances} \label{sect:3}
\hspace*{0.5cm}
In general, it is seen that resonances with wide energy-dependent widths
modify the dynamics of the processes in a non-trivial manner. Moreover, up
to now a definite way to obtain those widths directly from QCD is lacking.
The problem has been addressed in detail in Ref.~\cite{GomezDumm:2000fz},
where off-shell widths of resonances have been studied in the context of
R$\chi$T. In that work, vector meson resonances are analysed through the
two--point correlator of the vector current, defining the resonance
width as the imaginary part of the pole generated by the
resummation of loop diagrams that have absorptive contributions in the
s--channel. The widths obtained in this way are shown to satisfy the
requirements of analyticity, unitarity and chiral symmetry prescribed by
QCD. According to this definition, the energy-dependent width of the
$\rho(770)$ resonance is given by~\cite{GomezDumm:2000fz}~:
\begin{eqnarray}
\label{eq:rhowidth}
\Gamma_\rho(s) & = & \frac{M_{\rho} \, s}{96\,\pi\,F^2}
\,  \left[\sigma_\pi^3\,\theta(s\,-\,4m_\pi^2)\,+\,\frac{1}{2}\,
\sigma_K^3\,\theta(s\,-\,4m_K^2)\right]\,,
\end{eqnarray}
where $\s_P\,=\,\sqrt{1-4m_P^2/s}$. Incidentally it can be seen that an
analogous calculation for the $K^*(892)$ state leads to~:
\begin{eqnarray}
\label{eq:K*width} \Gamma_{K^*}(s) & = & \frac{M_{K^*}\,s}{128 \, \pi
F^2}\,\left[
\lambda^{3/2}\left(1,\,m_K^2/s,\,m_\pi^2/s\right)\,\theta
\left(s\,-\,(m_K\,+\,m_\pi)^2\right)\right.
\nonumber\\
& &\qquad \qquad \; \left.
+\,\lambda^{3/2}\left(1,\,m_K^2/s,\,m_\eta^2/s\right)\,\theta
\left(s\,-\,(m_K\,+\,m_\eta)^2\right)\right]\ ,
\end{eqnarray}
where $\lambda(a,b,c) = (a+b-c)^2-4ab$.
\par
In principle, one could apply the same definition in order to evaluate the
energy-dependent width of the a$_1$ resonance. However, this
involves a complex two-loop calculation within the resonance theory and
this is beyond our present reach. In Ref.~\cite{GomezDumm:2003ku} we
proposed an oversimplified approach in which the a$_1$ width was written in
terms of three parameters, namely the on-shell width
$\Gamma_{{\rm a}_1}(M_{{\rm a}_1}^2)$, the mass $M_{{\rm a}_1}$ and an exponent $\alpha$
that rules the asymptotic behaviour~:
\begin{equation}
\label{eq:a1width}
\Gamma_{{\rm a}_1}^{\mathrm{I}}(Q^2)\,=\,\Gamma_{{\rm a}_1}(M_{{\rm a}_1}^2)\,
\frac{\phi(Q^2)}{\phi(M_{{\rm a}_1}^2)}\,
\left(\frac{M_{{\rm a}_1}^2}{Q^2}\right)^\alpha\,\theta\left(Q^2\,-\,9
\,m_\pi^2\right)\, ,
\end{equation}
where
\begin{eqnarray}
\label{eq:phifunction}
\phi(Q^2) & = & Q^2\,\int\,\mathrm{d}s\;\mathrm{d}t\ \bigg\{
V_1^2\,|\mathrm{BW}_\rho(s)|^2\,+\,V_2^2\,|\mathrm{BW}_\rho(t)|^2
\nonumber\\
& & +\ 2\,(V_1\cdot V_2)\;{\rm Re}\left[
\mathrm{BW}_\rho(s)\,\mathrm{BW}_\rho(t)^*\right] \bigg\}\ .
\end{eqnarray}
Here $V_1$, $V_2$, $s$ and $t$ are defined as in the previous section,
the integral extends over the $3\pi$ phase space and the function
$BW_{\rho}(Q^2)$
is the usual Breit-Wigner for the $\rho(770)$ resonance, with the
energy-dependent width $\Gamma_\rho(q^2)$ given by
Eq.~(\ref{eq:rhowidth}).
\par
The analysis in Ref.~\cite{GomezDumm:2003ku}, which was previous to the
determination of the short-distance constraints from the three-point VAP
Green function~\cite{Cirigliano:2004ue}, left $\lambda_0$ unconstrained.
In this way, from the phenomenological analysis of experimental data on
$\tppp$ processes the following set of values was obtained: $\lambda_0
\simeq 12$, $\alpha \simeq 2.5$, $M_{{\rm a}_1} \simeq 1.2 \, \mbox{GeV}$ and
$\Gamma_{{\rm a}_1}(M_{{\rm a}_1}^2) \simeq 0.48 \, \mbox{GeV}$. On the other hand, if
one takes into account the relation in Eq.~(\ref{eq:lam3}), using $M_V =
M_{\rho(770)}$ and $M_A \simeq 1 \, \mbox{GeV}$ one gets $\lambda_0 \simeq 0.09$,
which differs drastically from the result quoted above (obtained from the
fit). In fact, as commented already in Ref.~\cite{GomezDumm:2003ku}, the
determination of $\lambda_0$ from the fit was undermined. The reason is
that in the $\tppp$ amplitude, without the constraint in
Eq.~(\ref{eq:lam3}), $\lambda_0$ appears always together with a
suppression factor $m_{\pi}^2/Q^2$ [see Eqs.~(\ref{eq:t1r}) and
(\ref{eq:fq2})], thus the dependence of the amplitude on this parameter
should be small. The result $\lambda_0\simeq 12$ is likely to be an
artifact to cure a wrong behaviour of the amplitude.
\par
In this work we stick to the short-distance constraints ruled by the VAP
Green function, thus we assume $\lambda_0$ as given by Eq.~(\ref{eq:lam3}). 
In fact, it will be seen that this value of
$\lambda_0$ is perfectly compatible with a proper description of $\tppp$
phenomenology. The new ingredients are the adoption of an adequate
definition for the energy-dependent width of the a$_1$ resonance,
and the inclusion of a small effect arising from the presence of a vector
resonance $\rho(1450)$ that we will consider in Sect.~\ref{sect:4}.
\par
We propose here a new parameterization of the a$_1$ width that is
compatible with the R$\chi$T framework used throughout our analysis. As
stated, to proceed as in the $\rho$ meson case, one faces the problem of
dealing with a resummation of two-loop diagrams in the two-point
correlator of axial-vector currents. However, it is still possible to
obtain a definite result by considering the correlator up to the two-loop
order only. The width can be defined in this way by calculating the
imaginary part of the diagrams through the well-known Cutkosky rules.
\par
\begin{figure}[t]
\begin{center}
\vspace*{0.9cm}
\includegraphics[scale=0.85,angle=0]{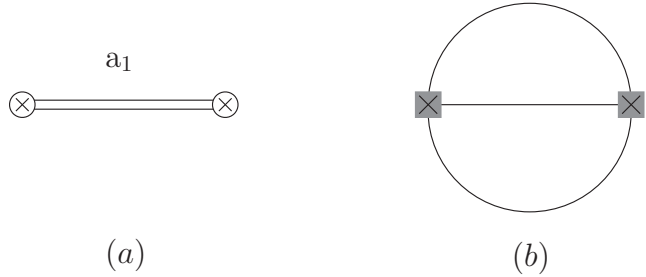}
\caption[]{\label{fig:2} \small{Diagrams contributing to the transverse part
of the correlator
of axial-vector currents in Eq.~(\ref{eq:correa2}). Diagram (a) gives
$\Pi_{(0)}^{{\rm a}_1}$ and diagram (b) provides $\Pi_{(1)}^{{\rm a}_1}$. The squared
axial-vector current insertion in (b) corresponds to the sum of the
diagrams in Fig.~\ref{fig:1}. The double line in (a) indicates the a$_1$ resonance
intermediate state. Solid lines in (b) indicate any Goldstone bosons
that carry the appropriate quantum numbers. }}
\end{center}
\end{figure}

Let us focus on the transversal component, $\Pi_T(Q^2)$, of the two-point Green function~:
\begin{eqnarray}\label{eq:correa1}
\Pi_{\mu\nu}^{33} & = & i \;\int\;{\rm d}^4x\; e^{iQ\cdot x}
\;\langle\, 0|\, T[A_\mu^3(x)\,A_\nu^3(0)]\, |0\, \rangle \nonumber \\
& = & (Q^2\,g_{\mu\nu}\,-\,Q_\mu Q_\nu)\;\Pi_{T}(Q^2) \, + \, Q_{\mu} Q_{\nu} \, \Pi_L(Q^2) \ ,
\end{eqnarray}
where $A_{\mu}^i = \overline{q} \gamma_{\mu} \gamma_5 \frac{\lambda^i}{2} q$. We will 
assume that the transversal contribution is dominated by the $\pi^0$ and the neutral component of the ${\rm a}_1$ triplet~: 
$\Pi_T(Q^2) \simeq \Pi^{\pi^0}(Q^2) +  \Pi^{{\rm a}_1}(Q^2)$.
Following an analogous procedure to the one in Ref.~\cite{GomezDumm:2000fz}, we
write $\Pi^{{\rm a}_1}(Q^2)$ as the sum
\begin{equation} \label{eq:correa2}
\Pi^{{\rm a}_1}(Q^2) \ = \ \Pi^{{\rm a}_1}_{(0)}\; + \;\Pi^{{\rm a}_1}_{(1)}\; +
\;\Pi^{{\rm a}_1}_{(2)}\; + \;\dots \ \ ,
\end{equation}
where $\Pi^{{\rm a}_1}_{(0)}$ corresponds to the tree level amplitude,
$\Pi^{{\rm a}_1}_{(1)}$ to a two-loop order contribution, $\Pi^{{\rm a}_1}_{(2)}$ to a
four-loop order contribution, etc. The diagrams to be included are those
which have an absorptive part in the $s$ channel. The first two terms are
represented by diagrams (a) and (b) in Fig.~\ref{fig:2}, respectively, where
effective vertices denoted by a square
correspond to the sum of the diagrams in Fig.~\ref{fig:1}. Solid lines
in the diagram (b) of Fig.~\ref{fig:2} correspond to any set
of light pseudoscalar mesons that carry the appropriate quantum numbers to
be an intermediate state.
\par
The first term of the expansion in Eq.~(\ref{eq:correa2}) arises from the
coupling driven by $F_A$ in the effective Lagrangian~(\ref{eq:ret}). We find
\begin{equation}
\Pi^{{\rm a}_1}_{(0)} \ = \ -\,\frac{F_A^2}{M_{{\rm a}_1}^2-Q^2} \ .
\label{pi0}
\end{equation}
Thus, if the series in Eq.~(\ref{eq:correa2}) can be resummed one should get
\begin{equation}
\Pi^{{\rm a}_1}(Q^2) \ = \ -\,\frac{F_A^2}{M_{{\rm a}_1}^2-Q^2 + \Delta(Q^2)} \ ,
\label{pitot}
\end{equation}
and the energy dependent width of the a$_1$ resonance can be defined by
\begin{equation}
M_{{\rm a}_1}\,\Gamma_{{\rm a}_1}(Q^2)\ = \ - \, {\rm Im}\,\Delta(Q^2) \ .
\end{equation}
Now if we expand $\Pi^{{\rm a}_1}(Q^2)$ in powers of $\Delta$ and compare term
by term with the expansion in Eq.~(\ref{eq:correa2}), from the second
term we obtain
\begin{equation}
\Delta(Q^2) \ = \
-\,\frac{(M^2_{{\rm a}_1}-Q^2)}{\Pi^{{\rm a}_1}_{(0)}}\ \Pi^{{\rm a}_1}_{(1)} \ .
\end{equation}
The off-shell width of the a$_1$ resonance will be given then by
\begin{equation}
\Gamma_{{\rm a}_1}(Q^2)\ = \
\,\frac{(M^2_{{\rm a}_1}-Q^2)}{M_{{\rm a}_1}\,\Pi^{{\rm a}_1}_{(0)}}
\ {\rm Im}\,\Pi^{{\rm a}_1}_{(1)} \ .
\label{gamma_pi1}
\end{equation}
\par
As stated, $\Pi^{{\rm a}_1}_{(1)}$ receives the contribution of various
intermediate states. These contributions can be calculated within our
theoretical R$\chi$T framework from the effective Lagrangian in
Eq.~(\ref{eq:ret}). In particular, for the intermediate $\pi^+\pi^-\pi^0$
state one has
\begin{eqnarray}
\Pi^{{\rm a}_1}_{(1)}(Q^2) & = & \frac{1}{6Q^2}\;\int\;
\frac{{\rm d}^4p_1}{(2\pi)^4}\;\frac{{\rm d}^4p_2}{(2\pi)^4}\
T^\mu_{1^+}\; T_{1^+\mu}^\ast\
\prod_{i=1}^3\; \frac{1}{p_i^2-m_\pi^2+i\epsilon} \ ,
\end{eqnarray}
where $p_3 = Q -p_1-p_2$, and $T_{1^+}$ is the $1^+$ piece of the hadron
tensor in Eq.~(\ref{tmu}),
\begin{equation}
T_{1^+}^{\mu} \; = \; V_1^\mu\,F_1 \, + \,  V_2^\mu\,F_2 \ .
\end{equation}
When extended to the complex plane, the function $\Pi^{{\rm a}_1}_{(1)}(z)$ has a
cut in the real axis for $z\geq 9m_\pi^2$, where ${\rm
Im}\,\Pi^{{\rm a}_1}_{(1)}(z)$ shows a discontinuity. The value of this imaginary
part on each side of the cut can be calculated according to the Cutkosky
rules as~:
\begin{equation}
{\rm Im}\,\Pi^{{\rm a}_1}_{(1)}(Q^2\pm i\epsilon) = \mp\frac{i}{2}\;\frac{1}{6Q^2}
\;\int\; \frac{{\rm d}^4p_1}{(2\pi)^4}\;\frac{{\rm d}^4p_2}{(2\pi)^4}\
T^\mu_{1^+}\; T_{1^+\mu}^\ast\ \prod_{i=1}^3\; (-2i\pi)\; \theta(p_i^0)\;
\delta(p_i^2-m_\pi^2) \ ,
\end{equation}
with $p_3 = Q -p_1-p_2$ and $Q^2 > 9 m_{\pi}^2$. After integration of the delta functions one finds
\begin{equation}
{\rm Im}\,\Pi^{{\rm a}_1}_{(1)}(Q^2\pm i\epsilon) = \pm\;\frac{1}{192\,Q^4}
\;\frac{1}{(2\pi)^3}\;\int\;{\rm d}s\,{\rm d}t \
T^\mu_{1^+}\; T_{1^+\mu}^\ast \ ,
\end{equation}
where the integrals extend over a three-pion phase space with total momentum
squared $Q^2$. Therefore, the contribution of the $\pi^+\pi^-\pi^0$ state to
the a$_1$ width will be given by
\begin{equation} \label{eq:Gamma_a1_pi}
\Gamma_{{\rm a}_1}^{\pi}(Q^2) \ = \ \frac{-1}{192(2\pi)^3 F_A^2 M_{{\rm a}_1}}\,
\left( \frac{M_{{\rm a}_1}^2}{Q^2}-1 \right)^2 \; \int
\mathrm{d}s\, \mathrm{d}t\; T^{\mu}_{1^+}\; T^{\ast}_{1^+\mu} \; .
\end{equation}
In the same way one can proceed to calculate the contribution of the
intermediate states $K^+K^-\pi^0$, $K^0\bar K^0\pi^0$, $K^-K^0\pi^+$ and
$K^+\bar K^0\pi^-$. The corresponding hadron tensors $T^K_{1^+}$ can be
obtained from Ref.~\cite{Dumm:2009kj}. Additionally one could consider the
contribution of $\eta \pi \pi$ and $\eta\eta\pi$ intermediate states.
However, these are suppressed by tiny branching ratios \cite{PDG2008,Gilman:1987my}
and will not be taken into account.
\par
In this way we have 
\footnote{It is important to stress that we do not intend to carry out the resummation of the
series in Eq.~(\ref{eq:correa2}). In fact, our expression in
Eq.~(\ref{gamma_pi1}) would correspond to the result of the resummation if
this series happens to be geometric, which in principle is not
guaranteed \cite{GomezDumm:2000fz}.}
\begin{equation} \label{eq:Gamma_a1_tot}
\Gamma_{{\rm a}_1}(Q^2) \ = \ \Gamma_{{\rm a}_1}^\pi(Q^2)\,  \theta(Q^2-9m_{\pi}^2) \
+ \ \Gamma_{{\rm a}_1}^K(Q^2) \, \theta(Q^2-(2 m_K+m_{\pi})^2)  \ ,
\end{equation}
where
\begin{equation} \label{eq:Gamma_a1_pimod}
\Gamma_{{\rm a}_1}^{\pi,K}(Q^2) \ = \ \frac{-S}{192(2\pi)^3 F_A^2 M_{{\rm a}_1}}\;
\left( \frac{M_{{\rm a}_1}^2}{Q^2}-1 \right)^2 \; \int \mathrm{d}s\, \mathrm{d}t\;
T^{\pi,K\mu}_{1^+}\; T^{\pi,K\ast}_{1^+\mu} \ .
\end{equation}
Here $\Gamma_{{\rm a}_1}^{\pi}(Q^2)$ recalls the three pion contributions and
$\Gamma_{{\rm a}_1}^K(Q^2)$ collects the contributions of the $KK\pi$ channels.
In Eq.~(\ref{eq:Gamma_a1_pimod}) the symmetry factor $S = 1/n!$ reminds
the case with $n$ identical particles in the final state. It is also important to point
out that, contrarily to the width we proposed in
Ref.~\cite{GomezDumm:2003ku} [$\Gamma_{{\rm a}_1}^{\mathrm{I}}(Q^2)$, in
Eq.~(\ref{eq:a1width})], the on-shell width $\Gamma_{{\rm a}_1}(M_{{\rm a}_1}^2)$ is
now a prediction and not a free parameter.
\par
An additional point to be taken into account are the off-shell widths of the
vector meson resonances entering the form factors in
$T_{1^+}^{\pi,K}$. Once again, since these resonances reach their on-shell
peaks in the phase space integrals, it is necessary to go beyond the leading
$1/N_C$ limit and include the corresponding energy-dependent widths. The
involved resonances for the $\pi\pi\pi$ and $KK\pi$ intermediate states
(always sticking to the approximation of taking only the lowest nonets) are
the $\rho(770)$, $K^\ast(892)$, $\omega(782)$ and $\phi(1020)$ vector
mesons. For the $\rho$ and $K^\ast$ we will consider the energy-dependent
widths in Eqs.~(\ref{eq:rhowidth}) and (\ref{eq:K*width}). Since resonances
$\omega(892)$ and $\phi(1020)$ are very narrow, the energy dependence is
irrelevant, and for our purposes we can take the experimental on-shell
widths quoted by the PDG~\cite{PDG2008}.

\section{The contribution of the $\rho(1450)$} \label{sect:4}
\begin{figure}[!t]
\begin{center}
\vspace*{0.9cm}
\includegraphics[scale=0.69,angle=-90]{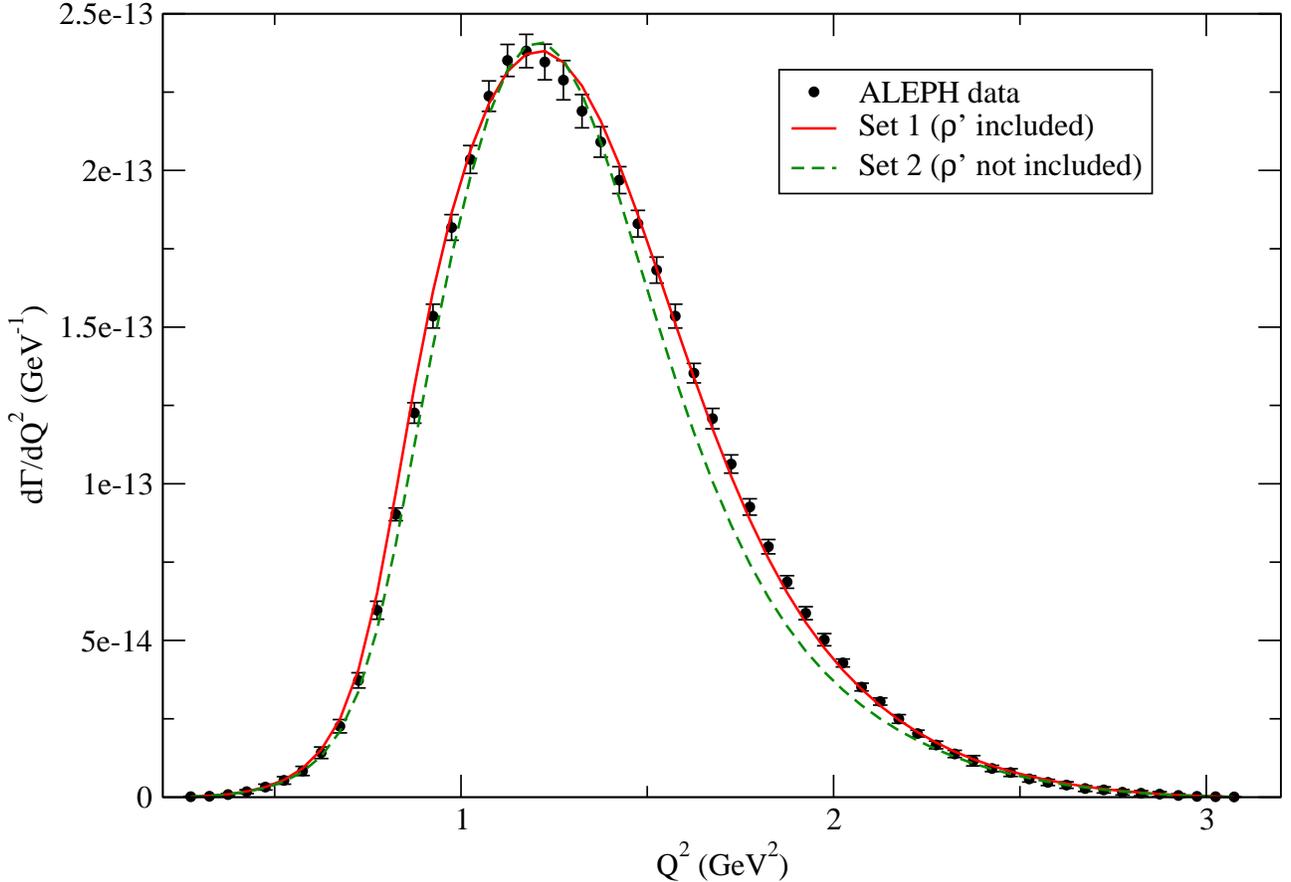}
\caption[]{\label{fig:3} \small{Comparison between the theoretical
$M_{3\pi}^2$-spectra of the $\tau^- \rightarrow \pi^+ \pi^- \pi^-
\nu_{\tau}$  with ALEPH data \cite{Barate:1998uf}. Set~1 corresponds to
the values of the parameters~: $F_V = 0.180 \, \mbox{GeV}$, $F_A = 0.149
\, \mbox{GeV}$, $M_{{\rm a}_1} = 1.120 \, \mbox{GeV}$, $\beta_{\rho'} = -0.25$,
$M_A \simeq 0.94 \, \mbox{GeV}$. Set~2 corresponds to the values of the
parameters~: $F_V = 0.206 \, \mbox{GeV}$, $F_A = 0.145 \, \mbox{GeV}$,
$M_{{\rm a}_1} = 1.150 \, \mbox{GeV}$, $\beta_{\rho'}=0$, i.e. without the
inclusion of the $\rho'$. In the case of Set 2 the overall normalization
of the spectrum has been corrected by a $15 \%$ to match the experimental
data.}}
\end{center}
\end{figure}

\hspace*{0.5cm} It turns out that, though some flexibility is allowed
around the predicted values for the parameters, the region between
$1.5-2.0 \, \mbox{GeV}^2$ of the three pion spectrum is still poorly
described by the scheme we have proposed here. This is not surprising as
the $\rho(1450)$, acknowledgeably rather wide, arises in that energy
region. We find that it is necessary to include, effectively, the role of
a $\rho' \equiv \rho(1450)$, in order to recover good agreement with the
experimental data. The $\rho'$ belongs to a second, heavier, multiplet of
vector resonances that we have not considered in our procedure. Its
inclusion would involve a complete new set of analogous operators to the
ones already present in ${\cal L}_{\rm R\chi T}$, Eq.~(\ref{eq:ret}), with
the corresponding new couplings. This is beyond the scope of our analysis.
However we propose to proceed by performing the following substitution in
the $\rho(770)$ propagator~:
\begin{equation}
 \frac{1}{M_{\rho}^2-q^2-iM_{\rho} \Gamma_{\rho}(q^2)} \longrightarrow
\frac{1}{1+\beta_{\rho'}} \, \left[ \frac{1}{M_{\rho}^2-q^2-iM_{\rho} \Gamma_{\rho}(q^2)} \, + \,
\frac{\beta_{\rho'}}{M_{\rho'}^2-q^2-iM_{\rho'} \Gamma_{\rho'}(q^2)} \right] \, ,
\end{equation}
where as a first approximation the $\rho'$ width is given by the decay into two pions~:
\begin{eqnarray}
 \Gamma_{\rho'}(q^2) & = &  \Gamma_{\rho'}(M_{\rho'}^2) \, \frac{M_{\rho'}}{\sqrt{q^2}}\,
\left( \frac{p(q^2)}{p(M_{\rho'}^2)} \right)^3 \, \theta ( q^2 - 4 \,m_{\pi}^2 ) \, , \\ \nonumber
p(x) & = &  \frac{1}{2} \, \sqrt{x-4 m_{\pi}^2} \, .
\end{eqnarray}
For the numerics we use the values $M_{\rho'} = 1.465 \, \mbox{GeV}$ and
$\Gamma_{\rho'}(M_{\rho'}^2) = 400 \, \mbox{MeV}$ as given in
Ref.~\cite{PDG2008}. We find that a
good agreement with the spectrum, $d \Gamma/dQ^2$, measured by ALEPH \cite{Barate:1998uf} 
is reached for the set of values~:
\begin{eqnarray} \label{eq:set1}
F_V \, = \, 0.180 \, \mbox{GeV} \; & \; , \;  & \;  F_A \, = \, 0.149 \, \mbox{GeV} \; \; \; \; \; \; \;  , \; \; \; \;
\beta_{\rho'} \, = -0.25  \;,  \nonumber \\
M_V \, = \, 0.775 \, \mbox{GeV}  \; & \; , \;  & \;
M_{K^*} \, = \, 0.8953 \, \mbox{GeV}\;  \; \; , \; \; \; \;
M_{{\rm a}_1} \, = \, 1.120 \, \mbox{GeV}  \, ,
\end{eqnarray}
that we call Set~1. The corresponding width is $\Gamma (\tau \rightarrow
\pi \pi \pi \nu_{\tau}) = 2.09 \, \times \, 10^{-13} \,\mbox{GeV}$, in
excellent agreement with the experimental figure $\Gamma (\tau \rightarrow \pi
\pi \pi \nu_{\tau})|_{exp} = (2.11 \pm 0.02) \, \times \, 10^{-13}
\,\mbox{GeV}$ \cite{PDG2008}. From $F_V$ and $F_A$ in Eq.~(\ref{eq:set1}), and the second
Weinberg sum rule we can also determine the value of $M_A = F_V M_V / F_A 
\simeq 0.94 \, \mbox{GeV}$, a result consistent with the one
obtained in Ref.~\cite{Mateu:2007tr}. If, instead, we do not include
the $\rho'$ contribution, the best agreement with experimental data is
reached for the values of Set~2~:
\begin{eqnarray} \label{eq:set2}
F_V \, = \, 0.206 \, \mbox{GeV} \; & \; , \;  & \;  F_A \, = \, 0.145 \, \mbox{GeV} \; \; \; \; \; \; \;  , \; \; \; \;
\beta_{\rho'} \, = 0  \;,  \nonumber \\
M_V \, = \, 0.775 \, \mbox{GeV}  \; & \; , \;  & \;
M_{K^*} \, = \, 0.8953 \, \mbox{GeV}\;  \; \; , \; \; \; \;
M_{{\rm a}_1} \, = \, 1.115 \, \mbox{GeV}  \, ,
\end{eqnarray}
though the branching ratio is off by $15 \%$.
A comparison between the results for the
$\tppp$ spectra obtained from Sets 1,2 and the data provided by ALEPH is
shown in Fig.~\ref{fig:3}. Notice that we have corrected the results provided by 
Set~2 by a normalization factor of 1.15 in order to compare the shapes of the spectra. 
Though it is difficult to assign an error to our numerical values, by
comparing Set~1 and Set~2 we consider that a $15\%$ should be on the safe
side. Notice, however, that the error appears to be much smaller in
the case of $M_{{\rm a}_1}$. 
\par
The value that we get for $M_{{\rm a}_1}=1.120 \, \mbox{GeV}$ differs from the one we got 
in Ref.~\cite{GomezDumm:2003ku}, namely $M_{{\rm a}_1}=1.203 (3) \, \mbox{GeV}$ (the 
error only includes the fit procedure). The disparity is mainly an outcome of the different off-shell
width of the a$_1$ that we introduce in this article and that we consider much more appropriate.
It has to be taken into account that our definition of $M_{{\rm a}_1}$ is the one given by 
Eq.~(\ref{eq:subst}) that constitutes and approach consistent with the features of our scheme. 
The physical pole mass could show a correction that, within our procedure, is a next-to-leading
effect in the $1/N_C$ expansion. It is also necessary to point out that masses and on-shell 
widths collected in the PDG \cite{PDG2008} also rely on models of the form factors or
scattering amplitudes.
\par
For Set 1 the width of the a$_1$ is $\Gamma_{{\rm a}_1}(M_{{\rm a}_1}^2) = 0.483
\, \mbox{GeV}$, which, incidentally, is in agreement with the figure we
got in Ref.~\cite{GomezDumm:2003ku} from a fit to the data. 
The value of $\Gamma_{{\rm a}_1}(M_{{\rm a}_1}^2)$ quoted in the PDG (2008)~\cite{PDG2008} 
goes from $250 \, \mbox{MeV}$ up to $600 \, \mbox{MeV}$.
\par
Our preferred set of values in Eq.~(\ref{eq:set1}) satisfies reasonably well
all the short distance constraints pointed out in Sect.~\ref{sect:2}, with a
deviation from Weinberg sum rules of at most $10 \%$, perfectly
compatible with deviations due to the single resonance approximation.

\section{Conclusions}
\hspace*{0.5cm}
The data available in $\tppp$ decays provide an excellent benchmark to study the hadronization of the
axial-vector current and, consequently, the properties of the a$_1(1260)$ resonance. In this article
we give a description of those decays within the framework of resonance chiral theory and the large-$N_C$
limit of QCD that: 1) Satisfies all constraints of the asymptotic behaviour, ruled by QCD, of the relevant
two and three point Green functions; 2) Provides an excellent description of the branching ratio and spectrum
of the $\tppp$ decays.
\par
Though this work was started in Ref.~\cite{GomezDumm:2003ku}, later
achievements showed that a deeper comprehension of the dynamics was needed
in order to enforce the available QCD constraints. To achieve a complete
description we have defined a new off-shell width for the a$_1$
resonance in Eq.~(\ref{eq:Gamma_a1_tot}), which is one of the main results
of this work. Moreover we have seen that the inclusion of the $\rho(1450)$
improves significantly the description of the observables. In
passing we have also obtained the mass value $M_{{\rm a}_1} = 1.120 \,
\mbox{GeV}$ and the on-shell width $\Gamma_{{\rm a}_1}(M_{{\rm a}_1}^2) = 0.483 \,
\mbox{GeV}$.
\par
With the description of the off-shell width obtained in this work we can now consider that the
hadronization of the axial-vector current within our scheme is complete and it can be applied in
other hadron channels of tau decays.

\paragraph{Acknowledgements} $\,$ \\ \label{Acknowledgements}
$\,$ \\
\hspace*{0.5cm} We wish to thank S.~Eidelman, M.~Jamin, O.~Shekhovtsova and Z.~Was for many useful
discussions on the topic of this article.
P.~Roig has been partially supported by a FPU contract (MEC), the DFG cluster of excellence
`Origin and Structure of the Universe' and a Marie Curie ESR Contract (FLAVIAnet).
This work has been supported in part by the EU MRTN-CT-2006-035482 (FLAVIAnet),
by MEC (Spain) under grant FPA2007-60323, by the Spanish Consolider-Ingenio 2010
Programme CPAN (CSD2007-00042) and by Generalitat Valenciana under grant PROMETEO/2008/069.
This work has also been supported by CONICET and ANPCyT (Argentina), under grants PIP6009,
PIP02495, PICT04-03-25374 and PICT07-03-00818.

\end{document}